\documentclass[aoas,MSNbibl,nameyear,noautosecdot,dvips]{arximspdf}
\usepackage{graphicx}
\doi{10.1214/13-AOAS667} 
\volume{7}
\issue{4}
\pubyear{2013}
\firstpage{1917}
\lastpage{1939}

\makeatletter

\renewcommand{\to}{\rightarrow}

\newcommand{\toProb}{\stackrel{p}{\rightarrow}}

\newcommand{\sub}{\subseteq}
\newcommand{\gv}{|}

\newcommand{\R}{\mathbb{R}}
\renewcommand{\P}{\mathbb{P}}
\newcommand{\E}{\mathbb{E}}
\newcommand{\D}{\mathcal{D}}

\newcommand{\IPP}{\mathrm{IPP}}

\newtheorem{proposition}{Proposition}

\makeatother

\begin{document}
\begin{frontmatter}

\title{Finite-sample equivalence in statistical models
for~presence-only data}
\runtitle{Statistical models for presence-only data}

\begin{aug}
\author[A]{\fnms{William} \snm{Fithian}\corref{}\thanksref{tWill}\ead[label=eWill]{wfithian@stanford.edu}}
\and
\author[A]{\fnms{Trevor} \snm{Hastie}\thanksref{tTrev}\ead[label=eTrev]{hastie@stanford.edu}}
\runauthor{W. Fithian and T. Hastie}
\affiliation{Stanford University}
\address[A]{Department of Statistics\\
Stanford University\\
390 Serra Mall\\
Stanford, California 94305-4065\\
USA\\
\printead{eWill}\\
\hphantom{E-mail: }\printead*{eTrev}} 
\end{aug}

\thankstext{tWill}{Supported by NSF VIGRE Grant DMS-0502385.}
\thankstext{tTrev}{Supported in part by NSF Grant DMS-10-07719 and NIH Grant
RO1-EB001988-15.}

\received{\smonth{10} \syear{2012}}
\revised{\smonth{6} \syear{2013}}

%
\begin{abstract}
Statistical modeling of presence-only data has attracted much recent
attention in the ecological literature, leading to a proliferation of
methods, including the inhomogeneous Poisson process (IPP) model,
maximum entropy (Maxent) modeling of species distributions and logistic
regression models. Several recent articles have shown the close
relationships between these methods. We explain why the IPP intensity
function is a more natural object of inference in presence-only studies
than occurrence probability (which is only defined with reference to
quadrat size), and why presence-only data only allows estimation of
relative, and not absolute intensity of species occurrence.

All three of the above techniques amount to parametric density
estimation under the same exponential family model (in the case of the
IPP, the fitted density is multiplied by the number of presence records
to obtain a fitted intensity). We show that IPP and Maxent give the
exact same estimate for this density, but logistic regression in
general yields a different estimate in finite samples. When the model
is misspecified---as it practically always is---logistic regression and
the IPP may have substantially different asymptotic limits with large
data sets. We propose ``infinitely weighted logistic regression,''
which is exactly equivalent to the IPP in finite samples. Consequently,
many already-implemented methods extending logistic regression can also
extend the Maxent and IPP models in directly analogous ways using this
technique.
\end{abstract}

%

\end{frontmatter}

\section{\texorpdfstring{Introduction.}{Introduction}}

In recent years ecologists have devoted significant attention to
the problem of estimating the geographic distribution
of a species of interest from records of where it has been found in the
past. There are many motivations for solving this problem, including
planning wildlife management actions, monitoring\vadjust{\goodbreak} endangered or
invasive species, and understanding species' response to different
habitats. A great variety of
experimental designs and statistical methods exist for tackling this
problem, and can be found in the literature on
resource-selection functions [\citet
{manly2002resource,lele2006weighted}], case-augmented designs
[\citet{lee2006fitting,dorazio2012predicting}] and site
occupancy modeling [\citet{mackenzie2006occupancy}].

Ecologists have proposed many statistical methods for modeling such
data, including the inhomogeneous Poisson process
(IPP) model [\citet{WartonIPP}], maximum entropy (Maxent) modeling of
species distributions
[\citet{Phillipsetal2004}, Phillips, Anderson and Scha\-pire (\citeyear{Phillipsetal2006}),
\citet{Phillipsetal2008}]
and the logistic regression model along with
its various generalizations such as GAM, MARS and boosted
regression trees [\citet{ESL}].
See \citet{elith2006novel} for
discussion and comparison of these and other methods in common use.

In recent years several articles have emerged detailing connections
between the three modeling methods above. Each method
takes as its input a presence-only data set along with a set of background
points consisting of a regular grid or random sample
of locations in some geographic region of interest. \citet{WartonIPP}
showed that logistic regression estimates converge to
the IPP estimate when the size of the presence-only data set is fixed
and the background sample grows infinitely large. \citet{AartsIPP}
additionally described a variety of models for
presence-only and other data sets whose likelihoods may all be derived from
the IPP likelihood. \citet{renner2013equivalence} further explore the
connection between Maxent and the IPP, taking up
the important issue of how we might check the IPPs modeling assumptions.

Our primary aim in writing this paper is to provide additional clarity
to this topic, recapitulating and
deriving the results in a unified framework and
extending them in several directions. We view all three major methods
as solutions to the same parametric density estimation problem.

\subsection{\texorpdfstring{Presence-only data.}{Presence-only data}}

Modeling of species distributions
is simplest and most convincing when the observations
of species presence are collected systematically.
In a typical design, a surveyor
visits a one-square-kilometer patch of land for one hour and records
how many specimens she discovers in that interval. The records of
unsuccessful surveys are called absence records, a mild misnomer since
ecologists recognize that specimens could be present but go
undetected. A data set reflecting presence or absence of a species in
each sampling unit is called presence--absence data.

\begin{figure}

\includegraphics{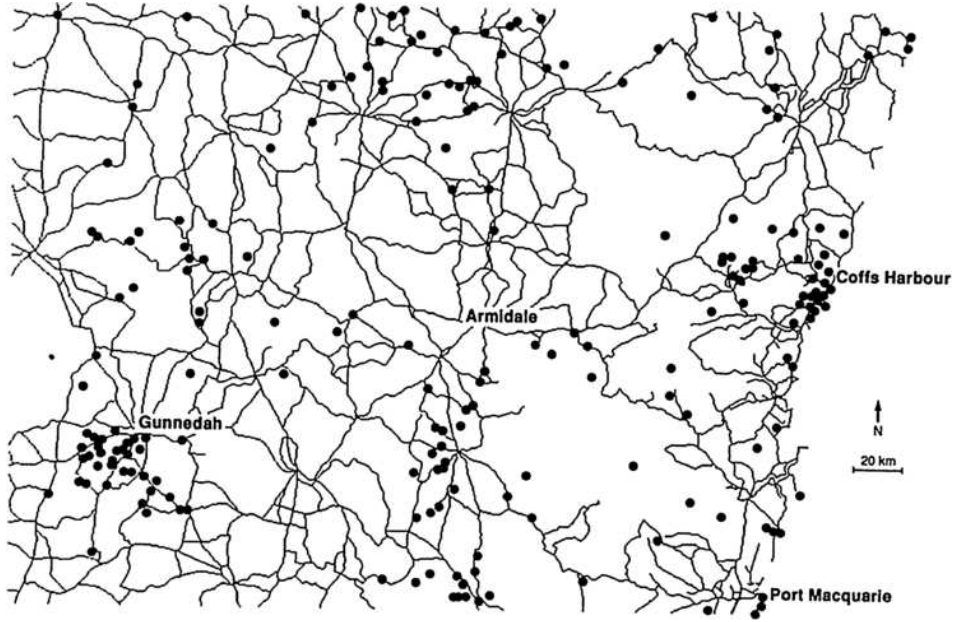}

\caption{Sampling bias in presence-only data for koalas.
Taken from Margules et~al. (\citeyear{margules1994biological}).}
\label{figRealKoala}\vspace*{-3pt}
\end{figure}

Unfortunately, dedicated surveys recording sampling effort are expensive,
especially for rare or elusive species. For many species of interest,
the only data available are museum or herbarium records of locations
where a specimen was\vadjust{\goodbreak} found and reported, for instance, by a motorist or
hiker. Typically these presence-only records are
collected haphazardly and frequently suffer from unknown
sampling bias such as that illustrated in Figure \ref{figRealKoala}.
The clustering of koala sightings near roads and cities probably has
more to do with the behavior of people than of koalas.

In recent years many such presence-only data sets have become
available electronically, and geographic information systems (GIS)
enable ecologists to remotely measure a variety of geographic
covariates without having to visit the actual locations of the
observations. As a result, presence-only data has
become a popular object of study in ecology [\citet{elith2006novel}].

\subsection{\texorpdfstring{What should we estimate\textup{?}}{What should we
estimate}}

Before we can sensibly decide how to model presence-only data, we
must address the issue of what it is we are modeling in the first
place. How should we think of ``species occurrence,''
the scientific phenomenon nominally under study? This issue arises
with presence-only and presence--absence data alike.

\subsubsection{\texorpdfstring{Occurrence probability.}{Occurrence
probability}}

\begin{figure}

\includegraphics{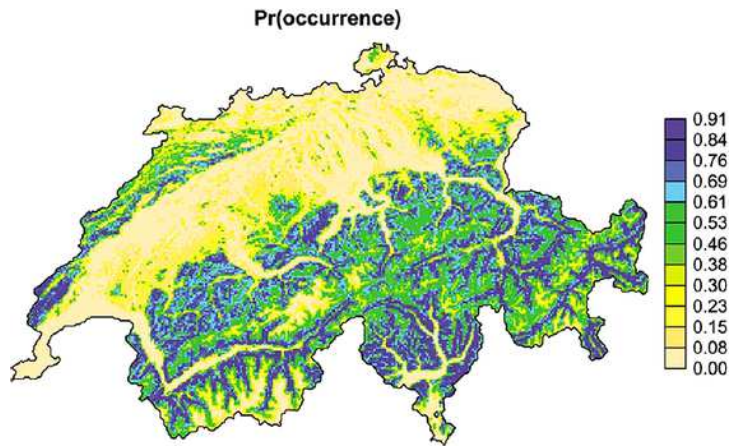}

\caption{Heat map of occurrence probabilities. Taken from
Royle, Nichols and K{\'{e}}ry (\citeyear{RoyleEtAlWillowTit}).}
\label{figWillowTit}
\end{figure}

Figure \ref{figWillowTit} is a typical ``heat-map'' output
of a study of the willow tit in
Switzerland using count data [\citet{RoyleEtAlWillowTit}]. The map reveals
which locations are more or less favored by the species (in this case,
high elevation and moderate forest cover appear to be the bird's
habitat of choice). The legend tells us
that the color of a region reflects the local probability of
``occurrence.''\vadjust{\goodbreak}

But precisely what event has this probability? Reading the
paper, we discover that occurrence means that
there is at least one willow tit present on a survey path through a 1
km${}\times{}$1 km quadrat of land. In this case, the authors analyze
a presence--absence data set using a hierarchical model that explicitly
accounts for the possibility that a bird was present but not detected
at the time of the survey.

Because the survey path length varies across sampling units, the
authors use it in their model as a predictor of presence probability.
It is not specified which
value of this predictor is used in generating the heat map, which
makes the map difficult to interpret.

Even if we could interpret the heat map as the probability of a bird
being present anywhere in the quadrat (not just along a path of
unspecified length), this probability would still be larger in a 2
km${}\times{}$2 km sampling unit and smaller in a 100~m${}\times{}$100
m one. Therefore, the very definition of ``occurrence probability'' in
a presence--absence study depends crucially on the specific sampling
scheme used to collect the presence--absence data. Consequently,
interpreting the legend of such a heat map can only make sense in the
context of a specific quadrat size, namely, whatever size was used in
the study. We would recommend that this information always be displayed
alongside the plot to avoid conveying the false impression (suggested
by a heat map) that occurrence probability is an intrinsic property of
the land, when it is really an extrinsic property.

Though the choice of quadrat size used to define occurrence
probability is ecologically arbitrary, it can in principle yield estimates
with meaningful interpretations.
By contrast, estimating occurrence probability in a
presence-only study is a murkier proposition. Any method purporting to
do so without reference to quadrat size
would be predicting the same occurrence probability within
a large or small quadrat, which cannot make sense.

\subsubsection{\texorpdfstring{Occurrence rate.}{Occurrence rate}}\label{subsecOccRate}

Since occurrence probability is only meaningful with reference to a
specific quadrat size, it is a somewhat awkward quantity to model in a
presence-only study. In this context it is more natural to estimate
an occurrence \emph{rate} or intensity: that is, a quantity with units of
inverse area (e.g., 1$/$km$^2$) corresponding to the expected number of
specimens \emph{per unit area}. Under some simple stochastic models
for species occurrence, including the Poisson process model considered
here, specifying
the occurrence rate is equivalent to specifying occurrence probability
simultaneously for all quadrat sizes.

Unfortunately, a presence-only data set only affords us direct knowledge
of the expected
number of specimen \emph{sightings} per unit area.
The absolute sightings rate is reflected in the
number of records in our data set, but, at
best, this rate is only proportional to the occurrence
rate discussed above, which typically is the real estimand of interest.
We must assume that our sightings only constitute a small fraction of
the species' population over our study region, possibly with repeated
sightings of the same specimen.
Without other data or assumptions we would have
no way of knowing what this constant of proportionality might be.

In other words, the absolute \emph{sightings} rate is observable but
usually not of direct interest, while the absolute \emph{occurrence}
rate is interesting but not observable without another source of
information. Using presence-only data alone, we can at best
hope to estimate a relative, not absolute, occurrence rate.
Even assuming that the sightings and occurrence rates are proportional
is optimistic, since it rules out sampling bias
like that in Figure \ref{figRealKoala}, an issue we take up again in
Section \ref{subsecObserverBias}.

\subsection{\texorpdfstring{Notation.}{Notation}}

We now introduce notation we will use for the remainder of the
article. We begin with some geographic domain of interest $\D$,
typically a bounded subset of $\R^2$. If the time of an observation
is an important variable, we might alternatively take
$\D\sub\R^3$, so that our observations have both space and
time coordinates. Associated with each geographic location $z\in\D$
is a vector $x(z)$ of measured features.

Our presence-only data set consists of $n_1$ locations of sightings
${z_i \in\D}$ for ${i=1,2,\ldots,n_1}$, accompanied by $n_0$
``background'' observations $z_i$ for $i=n_1+1,\ldots,n_1+n_0$
(typically a
regular grid or uniformly random sample from $\D$).
Finally, let $x_i=x(z_i)$ be the features for observation $i$,
and $y_i$ be a 0/1 indicator that $i$ is a presence sample.
Our treatment of these data as random or fixed will vary
throughout the article.

\subsection{\texorpdfstring{Outline.}{Outline}}

The rest of the paper is organized as follows. In
Section~\ref{secIPP} we define the log-linear
inhomogeneous Poisson process (IPP)\vadjust{\goodbreak} model
and its application to presence-only data, with special focus on
interpreting its parameters and their maximum likelihood
estimates.
In particular, the estimate of the intercept $\alpha$
reflects nothing more
than the total number of presence samples and, as such, is typically
not of scientific relevance for the reasons discussed in Section \ref
{subsecOccRate}. In fact,
IPP model estimation amounts to parametric density estimation in an
exponential family model, followed by multiplication of the fitted
density by
$n_1$. The density thus obtained
reflects the relative rate of sightings as a function of
geographic coordinates $z$.

\citet{AartsIPP} showed that many methods in species distribution
modeling can be
motivated by the IPP model. We review these connections and
generalize them for several illuminating examples.
In Section \ref{secMaxent} we consider a particularly important
example, showing
that the popular Maxent method of \citet{Phillipsetal2004} follows
immediately
from partially maximizing the IPP log-likelihood with respect to
$\alpha$, a result which is explored further in
\citet{renner2013equivalence}.
Hence, given any set of presence and background points,
the Maxent and IPP methods obtain identical estimates for the slope
$\hat\beta$ and for the density.

In Section \ref{secLogReg} we discuss so-called ``naive''
logistic regression and its connections to the IPP model. We derive
its likelihood as a conditional form of the IPP likelihood, but show
that if the
log-linear model is misspecified this convergence may
not occur until the background sample is quite large.
The need for a large
background sample is due not only to variance, but also to
bias that persists until the proportion $n_1/n_0$ becomes
negligibly small. We show, however, that if we upweight all the
background
samples by large weight $W\gg1$, we can
use logistic regression to recover the IPP estimate $\hat\beta$
precisely
with any finite presence and background sample. This
procedure, which we call ``infinitely weighted logistic regression,''
is a device for using GLM software to maximize the IPP
log-likelihood. Section \ref{secDiscussion} recapitulates the
relationships and
contains discussion.

\section{\texorpdfstring{The inhomogeneous Poisson process model.}{The inhomogeneous Poisson process model}}\label{secIPP}

The IPP is a simple model for a random
set of points $\mathbf{Z}$ falling in some domain $\D$. Both the
number and locations of points are random.
It can be defined by its intensity function
%
\begin{equation}
\lambda\dvtx  \D\longrightarrow[0,\infty),
\end{equation}
which indexes the likelihood that a point falls at or
near $z$. For $A\sub\D$, write
%
\begin{equation}
\Lambda(A) = \int_A{\lambda(z)\,dz}
\end{equation}
and assume $\Lambda(\D)<\infty$.

There are two main ways to formally characterize an IPP with
intensity $\lambda$. One simple definition is that
the total number of points is a Poisson random variable with
mean $\Lambda(\D)$ and, conditionally on the number of points,
their locations are independent
and identically distributed with density
$p_{\lambda}(z)=\lambda(z)/\Lambda(\D)$. That is, an IPP is an
i.i.d. sample from $p_\lambda$ whose size is itself
random.\setcounter{footnote}{2}\footnote{\citet{cressie1993} and \citet{AartsIPP} refer to an
IPP conditioned on $n_1$ as a ``Conditional IPP''; this is
exactly an i.i.d. sample of size $n_1$
from the density $p_{\lambda}(z)$.}

Alternatively, we can think of an IPP as a continuous limit of an
independent Poisson count model for ever-finer discretizations of $\D$.
If $N(A)=\#(\mathbf{Z}\cap A)$, the number of points falling in set
$A$, then
%
\begin{equation}
N(A) \sim\operatorname{Poisson}\bigl(\Lambda(A)\bigr)
\end{equation}
with $N(A)$ and $N(B)$ independent for disjoint sets $A$ and $B$. For
more on the IPP and other
point process models, see \citet{gaetan2009spatial} or \citet{cressie1993}.

In the case of a finite discrete
domain $\D=\{z_1,z_2,\ldots,z_m\}$, the IPP
model reduces to a discrete Poisson model, with
$N(z_i) \sim\operatorname{Poisson}(\lambda(z_i))$.
In this sense, the IPP model may be seen as a limit of finer and finer
discretizations of $\D$. We discuss this connection further in
Section \ref{subsecLLMConnect}.

\citet{WartonIPP}
proposed modeling species sightings
$z_1,\ldots,z_{n_1}$ as arising from an IPP whose
intensity is log-linear in the features $x(z)$:
%
\begin{equation}
\lambda(z) = e^{\alpha+ \beta'x(z)}.
\end{equation}
The formal
linearity assumption is less restrictive than it
seems, since our features $x(z)$ could include
polynomial terms, interactions, splines or other basis
expansions, which substantially broaden the space of possible
$\lambda(z)$.

Interpreting the IPP as an i.i.d. sample
with random size, we see that $\alpha$ and $\beta$ play very different
roles. Since $\alpha$ only multiplies $\lambda(z)$ by a constant, it
has no
effect on $p_\lambda(z)=\lambda(z)/\Lambda(\D)$. The ``slope''
parameters $\beta$
completely determine $p_\lambda$, while $\alpha$ scales the intensity
up or down to determine the expected sample size $\Lambda(\D)$.

\subsection{\texorpdfstring{Geographic space and feature space.}{Geographic space and feature space}}\label{subsecFeatSpace}

In the context of logistic regression,
it can be more natural to think
of the $x_i$ as a sample of points in ``feature space'' [i.e., the
range of $x(z)$] rather than
as the features corresponding to a sample in the geographic domain $\D$.
There is no real distinction between these two viewpoints, so long as
we adjust for the fact that
some values of $x$ are more common in $\D$ than others.

Let $A_x = \{z\dvtx  x(z) = x\}$ and $h(x) = \int_{A_x}1\,dz$. Then if
the set $\mathbf{Z}$ is an IPP with intensity
$\lambda(x(z))$, the corresponding set $x(\mathbf{Z})$ is an IPP with
intensity $\lambda_x(x) = \lambda(x)\cdot h(x)$ and, conditionally on
$n_1$, their distribution is $p_x(x) \propto p_\lambda(x)\cdot h(x)$.
For more detailed discussion see \citet{elith2011statistical} and
\citet{johnson2006resource}.

\subsection{\texorpdfstring{Maximum likelihood for the IPP.}{Maximum likelihood for the IPP}}\label{subsecMaxLikeIPP}

The score equations for the log-linear IPP are simple and
enlightening. The IPP log-likelihood in terms of the presence
samples is
%
\begin{equation}
\label{eqnIPPLoglikInt} \ell(\alpha,\beta) = \sum_{i:y_i=1}
\bigl(\alpha+\beta'x_i \bigr) - \int
_{\D}{e^{\alpha+\beta'x(z)}\,dz} - \log n_1!.
\end{equation}
Differentiating with respect to $\alpha$, we obtain the score equation
%
\begin{equation}
\label{eqnIPPScoreAlpha} n_1 = \int_\D{e^{\alpha+\beta'x(z)}
\,dz} = \Lambda(\D).
\end{equation}
That is, whatever $\hat\beta$ is, $\hat\alpha$ plays the role of a
``normalizing'' constant guaranteeing that $\lambda(z)$ integrates to
$n_1$, the number of total presence records.
Hence, if $n_1$ is not of scientific interest, then neither is
$\hat\alpha$.

Solving for $\alpha$ in (\ref{eqnIPPScoreAlpha}) and ignoring constants,
we obtain the partially maximized log-likelihood
%
\begin{equation}
\label{eqnIntCondLik} \ell^*(\beta) = \sum_{i:y_i=1}
\biggl(\beta'x_i - \log\int_\D{e^{\beta'x(z)}
\,dz} \biggr) = \sum_{i:y_i=1} \log p_\lambda(z_i),
\end{equation}
which is
the same log-likelihood we would obtain by conditioning on $n_1$ and
treating the $z_i$ as a random sample with density
$p_\lambda(z) = \frac{e^{\beta'x(z)}}{\int_\D
e^{\beta'x(z)}\,dz}$.\vspace*{1pt}

Finally, differentiating (\ref{eqnIntCondLik}) with respect to $\beta$
and dividing by $n_1$ gives the
remaining score equations:
%
\begin{equation}
\label{eqnBetaScore} \frac{1}{n_1}\sum_{i:y_i=1}x_i
= \frac{\int_\D{e^{\beta'x(z)}x(z)\,dz}} {
\int_{\D}{e^{\beta'x(z)}\,dz}} = \E_{p_\lambda} x(z).
\end{equation}
Solving (\ref{eqnBetaScore}) amounts to finding $\beta$ for which the
expectation of
$x(z)$ under $p_\lambda(z)$ matches the empirical mean over the presence
samples.

Hence, maximum likelihood for a log-linear IPP may be thought of as an
algorithm with two discrete steps:
\begin{enumerate}
\item Estimate the density $p_\lambda$:
find $\hat\beta$ for which $\E_{\hat p_\lambda}x(z)$ matches the
empirical means of the presence sample $x_i$.
\item Multiply $\hat p_\lambda$ by $n_1$: find $\hat\alpha$ for which
$\hat\lambda(z) =
n_1\cdot\hat p_{\lambda}(z)$.
\end{enumerate}
Unless $n_1$ is meaningful, then,
the IPP is essentially density estimation. In our view, it is
rare that $n_1$ merits much scientific interest, but there are
important cases where it might. For instance,
if we are comparing multiple species, study areas or periods of study,
and if we believe that sampling effort is comparable across the
different studies, then comparing the $n_1$ from each data set may
teach us something.

Note, however, that in each of these cases our
inference target can be viewed as a relative intensity \emph{across}
the different data sets.
If we wish to make such comparisons, the right approach may simply be to
expand the survey area $\D$ to include multiple regions or time
periods and add region identity or species identity as a feature, then
perform a combined analysis. $n_1$ for the combined analysis
(the total number of sightings across all the different data
sets) would then typically not be of much interest.

\subsection{\texorpdfstring{Numerical evaluation of the integral.}{Numerical evaluation of the integral}}\label{subsecNumeric}

When we cannot evaluate the integrals in
equations (\ref{eqnIPPLoglikInt})--(\ref{eqnBetaScore}) analytically,
we replace them with numerical integrals based on the background samples.
Hence, (\ref{eqnIPPLoglikInt}) becomes
%
\begin{equation}
\label{eqnIPPloglik} \ell(\alpha,\beta) = \sum_{i:y_i=1}
\alpha+\beta'x_i - \frac{|\D|}{n_0}\sum
_{i:y_i=0} e^{\alpha+\beta'x_i} - \log n_1!,
\end{equation}
where $|\D|=\int_{\D}{1\,dz}$ represents the total area of the region.

The background points may be either a uniform sample from $\D$ or a
regular grid. Quadrature weights may also be assigned to the
background points to approximate the integral with a weighted sum,
instead of the unweighted sum represented above.

We could repeat the derivation of Section \ref{subsecMaxLikeIPP} to
obtain the criteria
%
\begin{equation}
\label{eqnIPPScoreNumerical} \frac{|\D|}{n_0}\sum_{i:y_i=0}e^{\alpha
+\beta'x_i}
= n_1,\qquad \frac{ \sum_{i:y_i=0}e^{\beta'x_i}x_i} {
\sum_{i:y_i=0}e^{\beta'x_i}} = \frac{1}{n_1}\sum
_{i:y_i=1}x_i.
\end{equation}
Throughout, we will refer to (\ref{eqnIPPloglik}) as the numerical IPP
log-likelihood to distinguish it from
(\ref{eqnIPPLoglikInt}). In practice, fitting
the IPP means solving (\ref{eqnIPPScoreNumerical}) for some background
sample.

\subsection{\texorpdfstring{Connection to Poisson log-linear model.}{Connection to Poisson log-linear model}}\label{subsecLLMConnect}

If the background $z_i$
comprise a regular grid, we can discretize $\D$ into $n_0$
pixels $A_i$, each\vspace*{2pt} of roughly the same size $\frac{|\D|}{n_0}$
and centered at
$z_i$. If $x(z)$ is continuous, then
%
\begin{equation}
\label{eqnLLMapprox} \Lambda(A_i) = \int_{A_i}{e^{\alpha+\beta'x(z)}
\,dz} \approx\frac{|\D|}{n_0}e^{\alpha+\beta'x_i}.
\end{equation}
The IPP model implies that the counts $N(A_i)$
arise independently via 
%
\begin{equation}
N(A_i) \sim\operatorname{Poisson}\bigl(\Lambda(A_i)
\bigr) \approx\operatorname{Poisson} \biggl(\frac{|\D|}{n_0}e^{\alpha+\beta'x_i}
\biggr).
\end{equation}
Hence, the approximate log-likelihood is
%
\begin{eqnarray}
\label{eqnLLMLoglik} \tilde\ell(\alpha,\beta) &=& \sum
_{i:y_i=0}N(A_i) \bigl(\alpha+\beta'x_i
\bigr) -\frac{|\D|}{n_0}\sum_{i:y_i=0}e^{\alpha+\beta'x_i}\nonumber\\[-8pt]\\[-8pt]
&&{} -
\sum_{i:y_i=0}\log N(A_i)!.\nonumber
\end{eqnarray}
Let $S_i=\{k\dvtx z_k\in A_i,y_k=1\}$ contain the presence samples
in pixel $i$. Then
%
\begin{equation}
\sum_{i:y_i=0} N(A_i) \bigl(\alpha+
\beta'x_i\bigr) 
\approx\sum
_{i:y_i=0} \sum_{k\in S_i}\alpha+
\beta'x_k = \sum_{k:y_k=1}
\alpha+\beta'x_k.
\end{equation}
Hence, the only difference between (\ref{eqnIPPloglik}) and
(\ref{eqnLLMLoglik}) is that in the latter we also discretize the
location of each presence sample to match its nearest background point.

\citet{berman1992} proposed using this approximation to fit the IPP
model using Poisson GLM software, and \citet{baddeley2000practical}
show how to generalize it to other point-process models including
generalized additive models.
This device provides a simple means of
accessing the modeling flexibility of GLM methods
at a cost of some loss of data, since it effectively
replaces the covariate vector $x_i$ for
each presence sample with that of its nearest background sample.

\citet{baddeley2010spatial} discuss the bias incurred by the
discretization, showing in particular that it vanishes in the
small-pixel limit. They also propose a
strategy for improving the bias, which splits
pixels into subpixels
whose covariates are closer to constant.

As we will see later, this discretization is not really necessary. In
Section \ref{secLogReg} we propose a different procedure, infinitely
weighted logistic regression, that also
allows us to fit an IPP model using GLM software but produces exactly
the same estimates we would obtain by maximizing (\ref{eqnIPPloglik})
on the original presence and background data.

\subsection{\texorpdfstring{Identifiability and sampling bias.}{Identifiability and sampling bias}}\label{subsecObserverBias}

Sampling bias poses a serious challenge to valid inference
in presence-only studies. Scientifically, we are interested in the
\emph{occurrence process} consisting of all specimens of the species of
interest. However, our data set consists of what we might call
the \emph{sightings process},
consisting only of the occurrences observed and reported by people.


We can model the sightings process as an occurrence process
\emph{thinned} by incomplete observation, as proposed by
\citet{chakraborty2011point} and
\citet{renner2013equivalence}. That is, suppose that specimens
occur with intensity $\tilde\lambda(z)$, but that most occurrences go
unobserved. Each occurrence is observed with probability $s(z)$, which
may depend on features of the geographic location $z$ (e.g., proximity
to the road network). If detection is independent across occurrences,
then the observation process is an IPP with intensity
%
\begin{equation}
\lambda(z) = \tilde\lambda(z) \cdot s(z).
\end{equation}
The trouble is that our presence-only data set only
directly reflects $\lambda$, the intensity of sightings, and not
$\tilde\lambda$.


Optimistically, we might assume that $s$ is constant (no sampling bias).
In that case, by
estimating $\lambda(z)$ we are also
estimating
$\tilde\lambda(z)$ up to an unknown constant of
proportionality $s$, so\vadjust{\goodbreak} $p_{\tilde\lambda}=p_\lambda$ but
$\tilde\lambda\neq\lambda$. Even in this optimistic
scenario we can only
estimate relative, not absolute, occurrence intensities.
\citet{phillips2013estimating} also elaborate the same point
in the context of logistic regression models.


Slightly less optimistically, we might assume that $s$ is an unknown
function of~$z$, but that $s$ and $\tilde\lambda$ are known to
depend on $z$ through two disjoint feature sets. For instance,
we could model $\tilde\lambda$ and $s$ as log-linear in features
$x_1(z)$ and $x_2(z)$, respectively:
%
\begin{eqnarray}
\label{eqnLamOcc} \lambda(z) &=& \tilde\lambda(z)s(z)
\\
&=& e^{\tilde\alpha+\tilde\beta'x_1(z)}e^{\gamma+\delta'x_2(z)}.
\end{eqnarray}
Then the sightings process follows the log-linear model
$\lambda(z)=e^{\alpha+\beta'x(z)}$
with $\alpha=\tilde\alpha+\gamma$,
$x={x_1\choose x_2}$ and $\beta={\tilde\beta\choose\delta}$.
Note that $\tilde\alpha$ and $\tilde\beta$ are the quantities of
primary scientific interest, whereas $\alpha$ and $\beta$ are the
parameters governing the process we actually observe. Nevertheless,
$\tilde\beta$ is still identifiable from the data
because $\beta$ is.\footnote{As with any regression adjustment scheme,
we should proceed with caution here. If our linear
model is misspecified (perhaps we should have included
$x_2^2$)\vspace*{1pt}
and $x_1$ is correlated with the missing variables,
even regression adjustment will not remove all bias.
In perverse situations it could even make the situation worse.}

As $n_0,n_1\to\infty$, our estimate $\hat\beta_1$ converges to the
true value of $\tilde\beta$, the slope coefficients of
$\tilde\lambda$. However, $\hat\alpha$ will converge not to
$\tilde\alpha$ but rather to $\tilde\alpha+\gamma$. Without knowing
$\gamma$, we have no way of estimating $\tilde\alpha$. By the same
token, if
some features appear both in $x_1$ and $x_2$---or if $x_1$ and $x_2$
are not linearly independent---the model is
unidentifiable.

To be concrete, suppose koala occurrence is known to depend
only on elevation ($x_1$), and that sampling bias is known to depend
only on
proximity to roads~($x_2$). Then, despite the obvious sampling bias
in Figure \ref{figRealKoala}, we could still estimate what elevations koalas
tend to frequent, by making the correct adjustments for road proximity.
By contrast, we could not estimate from presence-only data alone whether
koalas tend to avoid roads, since that is confounded by
sampling bias.

Whether or not $s$ is constant,
our estimate for $\alpha=\tilde\alpha+\gamma$
carries no real information about $\tilde\alpha$ unless we have
independent knowledge of $\gamma$. Indeed, we have
already seen that the only role $\hat\alpha$
plays in estimation is to make $\lambda$ integrate to $n_1$.

The distinction between $\beta$ and $\tilde\beta$ may be very
important for some problems,
but for the remainder of this article we focus on estimation of $\beta$,
the slope parameters of the process we get to observe.


\section{\texorpdfstring{Maximum entropy.}{Maximum entropy}}\label{secMaxent}

Another popular approach to modeling presence-only data, which we
will see is equivalent to the IPP, is the
Maxent method proposed by \citet{Phillipsetal2004}.\vadjust{\goodbreak}
The authors begin by assuming that the presence samples
$z_1,\ldots,z_{n_1}$ are a random sample from some probability
distribution $p(z)$, called the species distribution.

The authors adopt the view, inspired by information theory, that our
estimate $\hat p$ should have large entropy
$H(p)=-\int_\D{p(z)\log(p(z))\,dz}$. Large $H(p)$ means roughly that
$p$ is close to the uniform density $1/|\D|$, the species distribution
we would observe if the species were indifferent to all geographic
features. The idea
is that $\hat p$ should be ``nearly geographically uniform,'' subject
to constraints that make it resemble the observed data.

\citet{Phillipsetal2004} propose to choose the $p$ which maximizes
$H(p)$ subject
to the constraint that the expectation of the features $x(z)$
under $\hat p$ matches the sample mean of those features, that is,
%
\begin{equation}
\label{eqnMeanConstr} \frac{1}{n_1}\sum_{y_i=1}x_i
= \int_\D{x(z)\hat p(z)\,dz} = \E_{\hat p} x(z).
\end{equation}
They show that this criterion is equivalent to
maximizing the likelihood of the parametric exponential family density:
%
\begin{equation}
\label{eqnMaxentloglik} p(z) = \frac{e^{\beta'x(z)}}{\int_\D{e^{\beta
'x(u)}\,du}}.
\end{equation}
This is exactly the form of $p_\lambda$ for our
log-linear IPP, and its log-likelihood is exactly the partially maximized
log-likelihood $\ell^*(\beta)$, the log-likelihood for
an IPP conditioned on $n_1$. The constraint (\ref{eqnMeanConstr}) is
precisely the score criterion (\ref{eqnBetaScore}) for $\beta$ in an
IPP, so the Maxent $\hat\beta$ is the same as the IPP $\hat\beta$.
This result may also be found in Appendix A of \citet{AartsIPP}.

The popular software package Maxent implements a method slightly more
complex than the one originally proposed in 2004. First, it automatically
generates a large
basis expansion of the original features into many derived features:
quadratic terms, interactions, step functions and hinge functions of
the original features. Then, it fits a model by optimizing
an $\ell_1$-regularized version of the conditional IPP
likelihood (\ref{eqnIntCondLik}):
%
\begin{equation}
\sum_{y_i=1}\beta'x_i -
n_1\log\biggl(\int_\D{e^{\beta'x(z)}\,dz}
\biggr) - \sum_j r_j|
\beta_j|.
\end{equation}
The regularization parameters $r_j$ are chosen automatically
according to rules based on an empirical study of various
presence-only data sets [\citet{Phillipsetal2008}].\footnote{The notation of the Maxent papers uses $\lambda$ and $\beta$
to denote what we call $\beta$ and $r$, respectively.}

Mathematically, the basis expansion increases the dimension of $x(z)$
but changes nothing else. Moreover, the $\ell_1$
regularization scheme does not constitute an essential difference with
the other methods considered here.\vadjust{\goodbreak}
One could (and often should) regularize $\beta$ when fitting an
IPP model as well, especially if $x(z)$
contains many features resulting from a large basis expansion.

Penalizing the Maxent log-likelihood does not
change the equivalence between the two models, so long as $\alpha$ is
left unpenalized.
If we add a penalty term $J(\beta)$ to the IPP log-likelihood
(\ref{eqnIPPLoglikInt}), we still obtain (\ref{eqnIPPScoreAlpha}) after
differentiating with respect to $\alpha$. Then,
partially maximizing $\ell(\alpha,\beta)-J(\beta)$ gives us
$\ell^*(\beta)-J(\beta)$, the penalized Maxent log-likelihood.
This equivalence depends on our not penalizing $\alpha$ in
(\ref{eqnIPPLoglikInt}).

This argument generalizes immediately to a generic penalized
likelihood method with any parametric form for $\log\lambda(z)$. We
have established the following general proposition:
%
\begin{proposition}\label{propIPPEquivCIPP}
Given some parametric family of real-valued functions
$\{f_\theta\dvtx  \theta\in\R^d\}$ with penalty function $J(\theta)$,
consider the penalized log-likelihood $g_1$ for an IPP with
intensity $e^{\alpha+f_\theta(x(z))}$,
%
\begin{equation}
\label{eqnGenIPP} g_1(\alpha,\theta) = \biggl(\sum
_{y_i=1}\alpha+f_\theta(x_i) \biggr) - \int
_\D{e^{\alpha+f_\theta(x(z))}\,dz} - J(\theta) - \log
n_1 !
\end{equation}
and the penalized log-likelihood $g_2$ for a sample
with density $\propto e^{f_\theta(x(z))}$:
%
\begin{equation}
\label{eqnGenCIPP} g_2(\theta) = \sum_{y_i=1}f_\theta(x_i)
- n_1\log\biggl(\int_\D{e^{f_\theta(x(z))}\,dz}
\biggr) - J(\theta).
\end{equation}
Then $\theta$ maximizes $g_2$ iff $(\alpha,\theta)$ maximize $g_1$
for some $\alpha$. The same applies if we
replace the integrals in (\ref{eqnGenIPP})--(\ref{eqnGenCIPP})
with sums over the background sample.
\end{proposition}

\begin{pf}
Partially maximize $g_1$ over $\alpha$ as in (\ref{eqnIntCondLik}) to
obtain $g_2$.
\end{pf}

Thus, we see that, while Maxent and the IPP appear to be different
models with different motivations, they result in the exact same density
estimate $\hat p_\lambda(z)$. In terms of the two-step
algorithm we derived in Section \ref{subsecMaxLikeIPP}, Maxent
is identical to step 1, but it skips step 2. The IPP fit
$\hat\lambda$ is $n_1$ times the Maxent fit $\hat p$.

\section{\texorpdfstring{Logistic regression.}{Logistic regression}}\label{secLogReg}

Another ostensibly different model for presence-only data is
so-called ``naive'' logistic regression, which casts
presence-only modeling as a problem of classifying points as presence
($y=1$) or background ($y=0$) on the basis of their features.
The logistic regression model
treats $n_1$, $n_0$ and the $x_i$ as fixed and the $y_i$ as
random with
%
\begin{equation}
\label{eqnLogReg} \P(y_i=1\gv x_i) = \frac{e^{\eta+\beta'x_i}}{1+e^{\eta
+\beta'x_i}}.
\end{equation}

Superficially, this approach may appear ad hoc and
unmotivated compared to IPP or Maxent. Nevertheless,
it has enjoyed some popularity,
in part because logistic regression
is an extremely mature method in statistics, enjoying myriad
well-understood and already-implemented
extensions such as GAM, MARS, LASSO, boosted regression
trees and more.

Logistic regression modeling of presence-only data has often been
motivated by analogy to logistic regression for presence--absence data.
Since it is not known whether the species is present at or near
the background examples, these are sometimes referred to as
``pseudo-absences,'' and the supposed naivete of the method is that it
appears to treat background samples
as actual absences. For instance, \citet{ward2009em}
introduced latent variables coding ``true'' presence or absence
and proposed fitting this model via the EM algorithm.

This interpretation raises once again the
troublesome question of what it would mean for one of our
randomly sampled background points to be a ``true presence.''
Need there be a specimen sitting directly
on the location, or is it enough for it to be within 100 m? 1 km?

Fortunately, we can sidestep these concerns, since
connections between the logistic regression and IPP
models yield a more straightforward interpretation.

%
%
%

\subsection{\texorpdfstring{Case-control sampling.}{Case-control sampling}}\label{subsecCCSamp}

Suppose the background data are a uniform random sample, and the
presence data arise from a log-linear IPP. Then if we condition on
$n_1$, the $z_i$ are a mixture of two i.i.d. samples,
one from density $e^{\alpha+\beta'x(z)}/\Lambda(\D)$ and the
other from density $1/|\D|$.
By Bayes' rule, for a random index $i$,
%
\begin{eqnarray}
\label{eqnBayes1} \P(y_i=1\gv z_i) &=& \frac{\P(y_i=1)\P(z_i\gv y_i=1)} {
\P(y_i=0)\P(z_i\gv y_i=0)+\P(y_i=1)\P(z_i\gv y_i=1)}
\\
&=& \frac{n_1e^{\alpha+\beta'x_i}/\Lambda(\D)} {
n_0/|\D|+n_1e^{\alpha+\beta'x_i}/\Lambda(\D)}
\\
\label{eqnBayes2} &=& \frac{e^{\eta+\beta'x_i}}{1+e^{\eta+\beta'x_i}},
\end{eqnarray}
with $e^\eta= \frac{n_1e^\alpha|\D|}{n_0\Lambda(\D)}$. Since\vspace*{1pt}
$\P(y_i=1\gv z_i)$ depends only on $x_i=x(z_i)$, we could just as well
condition on $x_i$ instead, giving (\ref{eqnLogReg}). Therefore,
if the log-linear IPP model is correct, it
implies the individual $y_i|x_i$
follow a logistic regression with the same slope parameters
$\beta$.\footnote{The $y_i$ are technically not conditionally
independent (if we knew the other $n_1+n_0-1$ labels, we would
know the last as well). This is always true in case-control studies,
but it is typically ignored since the dependence is weak for large
samples.}

Thus, given any finite sample of presence and background points, if we
believe in the IPP model, then we
could either maximize the numerical IPP likelihood or the
logistic regression likelihood, and in either case
we would be estimating the same population parameter $\beta$.
This does not guarantee we will obtain the same estimates
$\hat\beta$ in any given finite sample, but if the
model is correct, then either method gives a consistent estimator of
$\beta$.

Note that if we change
the marginal class ratio $n_1/n_0$ by some factor $e^c$,
the only effect will be to multiply the odds of $y_i=1$ given $x_i$ by
the same factor, that is, add $c$ to $\eta$ and leave $\beta$ unchanged.
Hence, under correct specification, $\hat\beta\to\beta$ regardless of
the limiting ratio $n_1/n_0$.

\subsection{\texorpdfstring{Case-control sampling under misspecification.}{Case-control sampling under misspecification}}\label{subsecMisspec}
Now, suppose that $\lambda(z)$ is
not really log-linear in our features $x$.
Then, the fitted slopes $\hat\beta$ for
logistic regression and the numerical
IPP will not converge to the same limiting
$\beta$ if $n_1$ and $n_0$ grow large together.
In fact, the limiting logistic
regression parameters depend on the limiting ratio of $n_1/n_0$
[\citet{xie1989logit}].

To gain some intuition for why this is so, suppose we have a single
covariate $x$, with $\lambda(z) =
e^{\alpha+x(z)^2}$. Then the derivation of
(\ref{eqnBayes1})--(\ref{eqnBayes2}) gives
%
\begin{equation}
\P(y_i=1\gv x_i) = \frac{e^{\eta+x_i^2}}{1+e^{\eta+x_i^2}}
\end{equation}
with $\eta$ as before.
In the large-sample limit, then, our estimation problem amounts to
finding $\hat\eta, \hat\beta$ for which
%
\begin{equation}
\label{eqnLogistApprox} \hat\eta+ \hat\beta x \approx\eta+ x^2
= \log\frac{n_1|\D|}{n_0\Lambda(\D)} + x^2
\end{equation}
in the population from which we are sampling. Now, since changing
$n_1/n_0$ only adds a vertical shift to the right-hand side of
(\ref{eqnLogistApprox}),
it may seem rather counterintuitive that this should
have any impact on the \emph{slope} $\hat\beta$ of our approximation on
the left-hand side.

To understand why,
we must come to grips with the sense in which
we make the approximation in (\ref{eqnLogistApprox}).
The logistic regression log-likelihood is
%
\begin{equation}
\ell_{\mathrm{LR}}(\eta,\beta) = \sum_{i} \bigl(
\eta+ \beta' x_i\bigr)y_i - \sum
_{i} \log\bigl(1+e^{\eta+\beta'x_i}\bigr).
\end{equation}
Its first derivatives with respect to $\eta$ and $\beta$ can be
written in terms of the fitted conditional probabilities
$\hat y_i(\eta,\beta) = \P_{\eta,\beta}(y=1\gv x=x_i)$:
%
\begin{eqnarray}
\frac{\partial\ell_{\mathrm{LR}}}{\partial\eta} &=& \sum_i
\biggl(y_i - \frac{e^{\eta+\beta'x_i}}{1+e^{\eta+\beta'x_i}} \biggr)
= \sum_i (y_i-\hat y_i),
\\
\frac{\partial\ell_{\mathrm{LR}}}{\partial\beta} &=& \sum_i x_i
\biggl(y_i-\frac{e^{\eta+\beta'x_i}}{1+e^{\eta+\beta
'x_i}} \biggr) = \sum
_i x_i(y_i-\hat
y_i).
\end{eqnarray}
If we define $r_i=y_i-\hat y_i$, then $\hat\eta, \hat\beta$
maximize the likelihood if and only if $\sum_i r_i = 0$ and $x \perp r$.
The crucial point is that the residuals of our approximation,
$y_i-\hat y_i$, are measured on the probability scale, and not the
log-odds scale.

\begin{figure}

\includegraphics{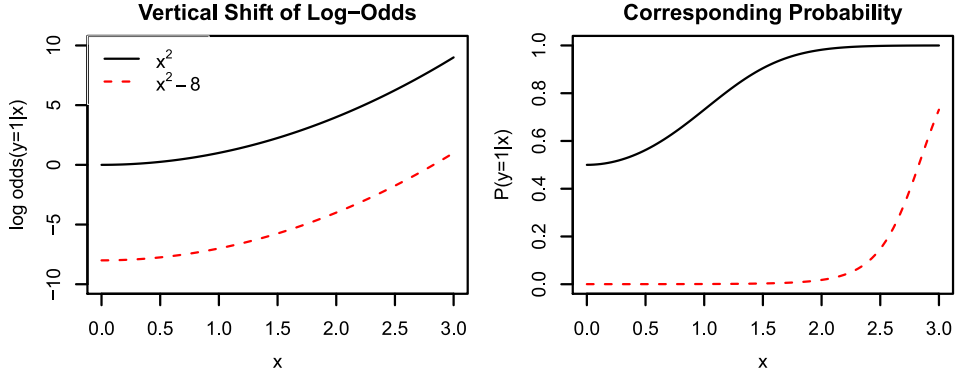}

\caption{The dashed red curve in the left panel is a vertical shift of
the solid black curve. However, vertically shifting the log-odds
changes the conditional probability in a more complex way.}
\label{figWarping}
\end{figure}

The black and red curves in the left panel of Figure \ref{figWarping}
show the
conditional log-odds $\log\frac{\P(y_i=1\gv x_i=x)}{\P(y_i=0\gv
x_i=x)}$ for our misspecified model with two different values of
$\eta$, 0 and $-8$, respectively. On the log-odds scale, one is no
steeper than the other. But when
we look at the same two curves on the conditional probability scale
(right panel), now the red looks steeper than the black.
This is due to a ``ceiling'' effect for the black curve:
in the region where the log-odds
$x^2$ is changing fast, the probability $\hat y =
\frac{e^{x^2}}{1+e^{x^2}}$ has already\vspace*{1pt} saturated at 1. The actual
estimates of $\hat\eta$ and $\hat\beta$ depend on the background
density of $x$ as well as $n_1/n_0$; see Section \ref{subsecSimStudy}
for a full simulation.

As \citet{WartonIPP} prove,\vspace*{1pt} this ceiling effect vanishes in the limit
where $n_1/n_0\to0$; in that case $\hat\eta\to-\infty$, $\hat y_i
= \frac{e^{\hat\eta+\hat\beta}}{1+e^{\hat\eta+\hat\beta}}
\approx e^{\hat\eta+ \hat\beta}$, and
the logistic\vspace*{1pt} regression and IPP estimates are identical.
Hence, there is no difference when the
background sample grows so large that it dwarfs the presence records
in the population from which we are sampling.
\citet{dorazio2012predicting} considers a similar framework, called the
case-augmented design, and proves a similar equivalency to the IPP as
$n_0\to\infty$.

\subsection{\texorpdfstring{Infinitely weighted logistic regression.}{Infinitely weighted logistic regression}}\label{subsecWtdLogReg}
If we modify the logistic regression procedure a
bit, we can resolve the discrepancy in the previous section
and recover the same $\hat\beta$ that we would estimate
with an IPP using the same presence and background samples.

We can remove the ceiling effect of the previous section if we add
case weights to the samples
%
\begin{equation}
w_i = \cases{W, &\quad $y_i=0$,
\cr
1, &\quad otherwise,}
\end{equation}
for some large number $W$. We then obtain the weighted
log-likelihood
%
\begin{eqnarray}
\label{eqnWLRloglik} \ell_{\mathrm{WLR}}(\eta,\beta) &=& \sum
_i w_i \bigl[y_i\bigl(\eta+
\beta'x_i\bigr) - \log\bigl(1+e^{\eta+\beta'x_i} \bigr)
\bigr]
\\
&=& \sum_{i:y_i=1}\eta+\beta'x_i
- \sum_{i}W^{1-y_i}\log
\bigl(1+e^{\eta+\beta'x_i} \bigr).
\end{eqnarray}

\begin{proposition}
Let $J(\beta)$ be any convex penalty, and suppose
$\ell_\IPP(\alpha,\beta)-J(\beta)$ has
a unique maximizer $(\hat\alpha_\IPP,\hat\beta_\IPP)$.
Then if $(\hat\eta_W,\hat\beta_W)$ maximize $\ell_{\mathrm
{WLR}}(\eta,\allowbreak\beta)-J(\beta)$ for
weight $W$,
%
\begin{equation}
\lim_{W\to\infty}\hat\beta_W = \hat\beta_\IPP.
\end{equation}
\end{proposition}

\begin{pf}
Reparameterizing (\ref{eqnWLRloglik}) with
$\alpha= \eta+ \log(W n_0/|\D|)$ and ignoring constants, we obtain
%
\begin{eqnarray}
\label{eqnWLRSub} \ell_{\mathrm{WLR}}(\alpha,\beta) &=& \sum
_{i:y_i=1}\alpha+\beta'x_i -\sum
_{i:y_i=0}W\log\biggl(1+\frac{|\D|}{Wn_0}e^{\alpha+\beta
'x_i} \biggr)
\nonumber\\[-8pt]\\[-8pt]
&&{} - \sum_{i:y_i=1}\log\biggl(1+
\frac{|\D|}{Wn_0}e^{\alpha+\beta
'x_i} \biggr).\nonumber
\end{eqnarray}
Fixing\vspace*{1pt} $(\alpha,\beta)$ and taking $W\to\infty$, each
term in the second sum converges to
$\frac{|\D|}{n_0}e^{\alpha+\beta'x_i}$ while the third sum
converges to
0. Hence, ignoring constants,
(\ref{eqnWLRSub}) converges to the numerical IPP
log-likelihood (\ref{eqnIPPloglik}), and this convergence occurs
uniformly on compact subsets of the parameter space.

Now, both $\ell_{\mathrm{WLR}}(\alpha,\beta)-J(\beta)$ and
$\ell_\IPP(\alpha,\beta)-J(\beta)$ are concave, and the
latter is strictly concave by assumption; hence, the maximizer of
the first converges to the maximizer of the second.
\end{pf}

From the above, we see that IWLR is not really a new statistical
method, but
rather a technical device for optimizing the IPP/Maxent log-likelihood
using already-implemented GLM software.

Although technically $\hat\beta_W\neq\hat\beta_\IPP$ for
any finite $W$ (hence the name ``infinitely weighted''), in practice,
we only need $W$ large enough that the approximation of
$\ell_{\mathrm{WLR}}(\alpha,\beta)$ to
$\ell_{\mathrm{IPP}}(\alpha,\beta)$ is good near
$(\hat\alpha,\hat\beta)$.\vspace*{2pt}

Essentially,\vspace*{1pt} if $\frac{|\D|}{Wn_0}e^{\alpha+\beta'x_i}\approx0$ for
each $i$ (say, all are less than 0.001), then the Taylor
approximation should be good. We can assess this easily if we observe
that
%
\begin{equation}
\hat y_i = \frac{{|\D|}e^{\hat\alpha+\hat\beta'x_i}/{(Wn_0)}} {
1+{|\D|}e^{\hat\alpha+\hat\beta'x_i}/({Wn_0})} \approx\frac{|\D
|}{Wn_0}e^{\hat\alpha+\hat\beta'x_i},
\end{equation}
when all of the above are small. To rephrase, then, if $\max_i \hat
y_i$ from the logistic regression is less than 0.001 or
so, it seems to us that $W$
should be sufficiently large. If not,\vspace*{1pt} we can set $W \gets
\frac{\max_i \hat y_i}{0.001} W$ and check that the fitted $\hat y_i$
are now small enough. If any uncertainty remains whether $W$ is large
enough, one can always increase it by (say) another factor of 100 and
check that the estimates do not change appreciably.

\subsection{\texorpdfstring{Logistic regression as density estimation.}{Logistic regression as density
estimation}}

One interpretation of the results we have just reviewed is that
in the context of presence-only data, logistic regression solves the
same parametric density
estimation problem as Maxent and the IPP do. Moreover, our
infinitely weighted
logistic regression yields an identical estimate of the density.

Using logistic regression for density estimation has been proposed
before. For example, \citet{ESL}
discuss it as a means for
turning an unsupervised density estimation problem into a supervised
classification problem. Their
proposal uses a different weighting scheme (assigning
half the total weight to the presence samples) which, unlike
infinitely weighted logistic regression,
does not give exactly the IPP solution.

\subsection{\texorpdfstring{Simulation study: Weighted vs unweighted logistic
regression.}{Simulation study: Weighted vs unweighted logistic
regression}}\label{subsecSimStudy}

We have seen that both infinitely weighted logistic regression
(a.k.a. numerical IPP) and unweighted
logistic regression estimate the same $\beta$ parameter of the same
log-linear IPP model, and when the background sample is much larger than
the presence sample the estimates $\hat\beta$ are close to each
other.

However, the infinitely weighted logistic regression estimate can
converge much faster to the large-background-sample limit if the linear
model is misspecified, as we illustrate here with a simulation study.

Consider a geographic region with a single covariate $x$ whose
background density is $p_0(x)=N(0,1)$. Now,\vspace*{1pt} suppose a species
follows our log-linear IPP model with slope $\beta$, so that
$\lambda(x(z))\propto e^{\beta x}$. Then the density of presence samples
in feature space is $p_1(x)=e^{\beta x}p_0(x)/
(\int{e^{\beta u}p_0(u)\,du} ) = N(\beta,1)$.

\begin{figure}

\includegraphics{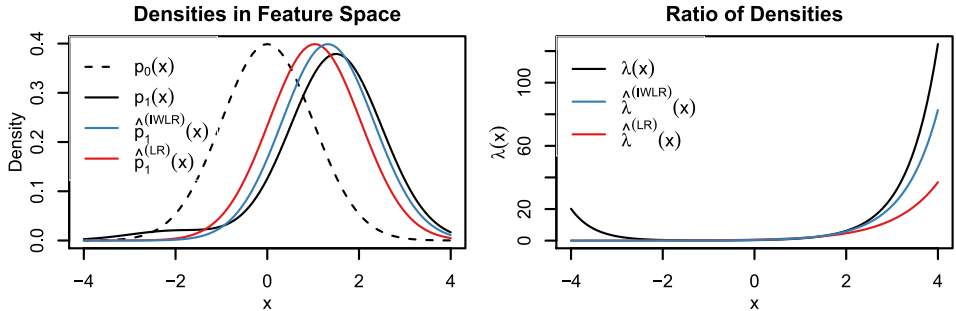}

\caption{Large-sample estimates for the simulation study,
misspecified case. The black curves represent the true presence
density (left panel) and intensity (right panel). The blue and
red curves show the fitted densities using IWLR and standard logistic
regression with $n_0=n_1$.}
\label{figSimApprox}
\end{figure}

Suppose our species is in fact a mixture of two subspecies, one of which
comprises 95\% of the population and prefers $x$ large, while the
remaining 5\% prefer $x$ small. If each subspecies
follows our model with coefficients 1.5 and $-2$, respectively, then
%
\begin{equation}
\lambda(x) \propto0.95 e^{1.5x} + 0.05 e^{-2x},
\end{equation}
which no longer follows the log-linear model. $p_0(x)$ and $p_1(x)$
are depicted in the upper panel of Figure \ref{figSimApprox} as the
dashed and solid black lines. The black line in the left panel shows
$\lambda(x)=p_1(x)/p_0(x)$, the relative intensity as a function of
the covariate $x$. In the left panel all the curves have been
normalized so that $\Lambda(\D)=\int\lambda(x)p_0(x)\,dx=1$.

If we\vspace*{1pt} fit an infinitely-weighted logistic regression (or, equivalently,
a log-linear IPP) to a large presence and background sample, our
fitted $\hat\beta^{(\mathrm{IWLR})}$ will tend to
$\mu_1=\E_{p_1}(x)=1.325$. We have plotted the corresponding
large-sample estimates
$\hat\lambda^{(\mathrm{IWLR})}(x)$ and $\hat p_1^{(\mathrm{IWLR})}(x)$
as blue lines in the respective panels
of Figure \ref{figSimApprox}.

If, alternatively,\vspace*{1pt} we fit an unweighted logistic
regression to the same data set with large $n_0=n_1$,
the estimate $\hat\beta^{(\mathrm{LR})}$ will tend to roughly
1.04. The resulting large-sample estimates
$\hat p_1^{(\mathrm{LR})}(x)$ and $\hat\lambda^{(\mathrm{LR})}(x)$ are
plotted in red. 

If we fit an unweighted logistic regression to a large sample with a
different ratio $n_1/n_0$, we would get a different estimate, which
would tend toward the IPP estimate of 1.325 if and only if this ratio
tended to 0. By the same token, when $n_1$ and $n_0$ are
fixed, the ratio between them can play a significant role in determining
the estimated $\beta$.
In contrast, the IWLR/IPP estimate tends to
1.325 in large samples no matter what the ratio $n_1/n_0$.

\begin{figure}

\includegraphics{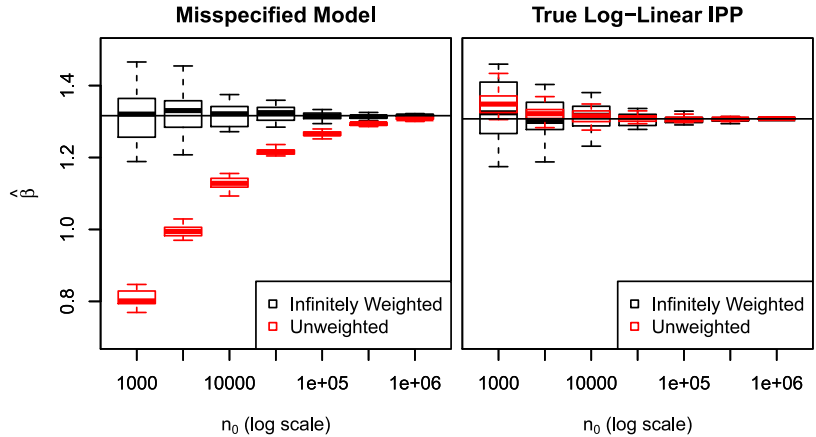}

\caption{$\hat\beta$ estimates for simulation study with $n_1=3000$
and varying $n_0$.
Unweighted logistic regression may require a very large
background sample before convergence when the model is misspecified.}
\label{figWtdUnwtdBox}
\end{figure}

The left panel of Figure \ref{figWtdUnwtdBox} illustrates this with a
simulation study of the example just discussed. We first generate a
single presence sample of size $n_1=3000$ from this species, then
generate 20 sets of $n_0$ background samples from $p_0=N(0,1)$ for each
of a
range of values $n_0$ ranging from $10^3$ to $10^6$.

For each background sample, we fit both an ``infinitely'' weighted
($W=10^4$) and
unweighted logistic regression to the combination of presence and
background points. For relatively
large sizes of background sample, there is very little sampling
variability, but the logistic regression estimates carry a large\vspace*{1pt} bias
that depends greatly on the size of the background sample. The
limiting $\hat\beta$, to which both methods would converge given an
infinite background sample, is depicted with a horizontal line.

In the right panel, we repeat this study with a presence sample
from $N(\mu_1,1)$, the correctly-specified model with the same
mean as our misspecified model. Now the situation is very different;
no matter what the mix of presence and background samples, the
log-odds are truly linear with slope $\beta=\mu_1$. Consequently,
$\hat\beta^{(\mathrm{LR})}\toProb\beta$ as $n_0\to\infty$ and
$n_1\to\infty$, regardless of the limiting
ratio $n_1/n_0$.

Since the choice of background sample size is primarily a matter of
convenience, it is preferable to use an estimator
that depends on it as little as possible. When the linear model is
misspecified (which is nearly always the case), we recommend the
infinitely weighted logistic regression over unweighted logistic
regression for this reason.

We emphasize here that although IWLR resolves the issue of
bias that we discussed in Section \ref{subsecMisspec}, using
IWLR does \emph{not} guarantee that we will obtain a good estimate for small
$n_0$. The smaller $n_0$ is, the larger
the variance of our estimate, so a larger background set is always
better unless computational constraints apply.

What is more, the variability in our
estimate due to the background sample is not reflected in the
default standard error outputs from GLM software---only the variability
due to the presence records is. Because\break
$\ell_{\mathrm{IWLR}}(\alpha,\beta)\approx
\ell_{\mathrm{IPP}}(\alpha,\beta)$
for large $W$, its Hessian will also converge to the Hessian
of the IPP.

Even if our background sample was extremely large, the standard error
estimates for any of the models we have discussed are based on
asymptotic normal approximations that hold when the log-linear model
is correctly specified. Resampling methods such as the bootstrap are
more generally reliable, but even\vadjust{\goodbreak} the bootstrap will depend crucially
on the
assumption that presence records (and in the case of logistic
regression, background records) are \emph{independent} observations.
In terms of the IPP model, this assumption rules out spatial
clustering of presence records.
\citet{renner2013equivalence} provide evidence that this assumption may
not hold for presence-only data. Therefore, model-based
estimates of standard error should be viewed with
suspicion no matter what method we choose.

\section{\texorpdfstring{Discussion.}{Discussion}}\label{secDiscussion}

We have discussed several closely related models for a single
presence-only sample. In this section we collect them all in one
place and review their relationships:

\begin{longlist}
\item[\textit{Inhomogeneous Poisson process}.]
The ``mother'' model, from which the others can be
derived, is the inhomogeneous Poisson process (IPP), whose
log-likelihood is
%
\begin{equation}
\label{eqnRevIPP} \sum_{i:y_i=1} \bigl(\alpha+
\beta'x_i \bigr) - \int_\D{e^{\alpha+\beta'x(z)}
\,dz}.
\end{equation}
In practice, (\ref{eqnRevIPP}) is approximated numerically via
%
\begin{equation}
\label{eqnRevIPPNum} \sum_{i:y_i=1} \bigl(\alpha+
\beta'x_i \bigr) - \frac{|\D|}{n_0}\sum
_{i:y_i=0} e^{\alpha+\beta'x_i}.
\end{equation}
Fitting\vspace*{1pt} this model amounts to solving for the density $p_\lambda(z)
\propto e^{\beta'x(z)}$ for which the expected features
$\E_{p_\lambda} x(z)$ match the empirical mean
$\frac{1}{n_1}\sum_{i:y_i=1}{x_i}$, then multiplying that density by $n_1$.

\item[\textit{Maxent}.]
Conditioning\vspace*{1pt} on $n_1$, we obtain the exponential family density model
$p(z) \propto e^{\beta'x(z)}$, resulting in the log-likelihood
%
\begin{equation}
\sum_{i:y_i=1}\beta'x_i -
n_1\log\biggl(\int_\D{e^{\beta'x(z)}\,dz}
\biggr)
\end{equation}
or its numerical counterpart. This is the log-likelihood maximized by
Maxent, and it corresponds exactly
to the log-likelihood (\ref{eqnRevIPP}) partially maximized with
respect to $\alpha$.
Hence, both procedures give exactly the same estimates of $\beta$ and $p$.

\item[\textit{Logistic regression}.]
The logistic regression log-likelihood is
%
\begin{equation}
\label{eqnRevLogReg} \sum_i y_i
\bigl(\eta+\beta'x_i\bigr) - \log\bigl(1+e^{\eta+\beta'x_i}
\bigr).
\end{equation}
When the log-linear IPP model is correctly specified, this model is as
well (aside from the fact that the $y_i|x_i$ are only approximately
independent), with the same true $\beta$ as in the IPP model.
However, in finite samples the estimates for $\beta$ given by
maximizing (\ref{eqnRevLogReg}) instead of (\ref{eqnRevIPPNum}) may be
substantially different.

\item[\textit{Infinitely weighted logistic regression}.]
We can resolve this difference by upweighting all the
background points by $W\gg1$, obtaining\vadjust{\goodbreak} weighted log-likelihood
%
\begin{equation}
\label{eqnRevWLR} \sum_{i:y_i=1} \bigl(\eta+
\beta'x_i \bigr) - \sum_{i}W^{1-y_i}
\log\bigl(1+e^{\eta+\beta'x_i} \bigr).
\end{equation}
In the limit where $W\to\infty$, we recover exactly the
same $\hat\beta$ as we would by maximizing (\ref{eqnRevIPPNum}).

\item[\textit{Discretized Poisson LLM}.]
Another means for approximating the IPP log-likelihood with a GLM
log-likelihood is the Berman and Turner method, which simply
discretizes geographic space into pixels and assigns each presence
point to a bin belonging to its nearest background point:
%
\begin{equation}
\label{eqnRevLLM} \sum_{i:y_i=0} N(A_i)
\bigl(\alpha+\beta'x_i\bigr) -\frac{1}{n_0}\sum
_{i:y_i=0}e^{\alpha+\beta'x_i}.
\end{equation}
This discretization of presence features
is unnecessary given that we can exactly fit the IPP
likelihood using the infinitely weighted approach of (\ref{eqnRevWLR}).
\end{longlist}


\subsection{\texorpdfstring{Extending the IPP model.}{Extending the IPP
model}}

Logistic regression is one of the most widely applied methods in
statistics. For decades, applied statisticians have been developing,
studying and using variations on logistic regression to solve
classification problems in statistics. R packages exist for fitting
generalized additive models (GAMs), boosted regression trees, MARS
and every manner of tailored regularization schemes [see, e.g.,
\citet{ESL}].

All of these methods are well understood within the
context of logistic regression. We believe that the most
important practical implication of the finite-sample equivalence
between the IPP model and infinitely weighted logistic regression is that
all of these methods can now be equally well understood and easily
applied within the context of the IPP model.

For instance, we can fit an IPP / Maxent version of
boosted regression trees with the following single line of R:
\begin{verbatim}
boosted.ipp <- gbm(y~., family=``bernoulli,''
data=dat, weights=1E3^(1-y)).
\end{verbatim}
For an IPP / Maxent version of LASSO, ridge, or the elastic
net:\footnote{The user should be warned that \texttt{glmnet}
automatically re-normalizes the weights so they sum to $n_0+n_1$.
To avoid issues, set \texttt{glmnet.control(pmin=1.0e-8,} \texttt{fdev=0)}
in your R session, and keep in mind this
renormalization when setting the Lagrange parameter $\lambda$.}
\begin{verbatim}
lasso.ipp <- glmnet(dat.x, dat.y, family=``binomial,''
weights=1E3^(1-y)).
\end{verbatim}
For an IPP GAM:
\begin{verbatim}
gam.ipp <- gam(y~s(x1)+x2, family=binomial, data=dat,
weights=1E3^(1-y)).
\end{verbatim}

This added flexibility promises to provide a powerful tool to modelers
of presence-only data.

\subsection{\texorpdfstring{Model selection.}{Model selection}}

Regardless of which of the various related likelihoods we choose, there
remains the issue of model selection. With the use of geographic
information
systems, ecologists often have access to a large number of predictor
variables and may wish to winnow the field before modeling to avoid
overfitting. Conversely, if some continuous variables are
known to be important predictors, assuming a linear effect on the
log-intensity may be too restrictive, and we may wish to expand the
basis using splines, interactions, wavelets, etc. In either case,
regularization may be called for.

Though it would be impossible to give a full treatment here of the
many important considerations governing model selection, we note that
these choices need not be governed by which likelihood
we take as our starting point. In particular, the
large set of derived features and $\ell_1$ regularization used by
Maxent software can just as well be applied to the IPP model or, for
that matter, to logistic regression. Using
the infinitely weighted logistic regression method, we can
implement the exact loss function used by the Maxent with software for
penalized GLMs.

\section*{\texorpdfstring{Acknowledgments.}{Acknowledgments}}

The authors are grateful to Jane Elith for helpful discussions and
suggestions. We were also very fortunate to have had very thorough and
thoughtful reviewers; the current manuscript has benefited greatly from
their attention.



\printaddresses

\end{document}